\documentclass[twocolumn,showpacs,amsmath,amssymb,prb]{revtex4}% 
\usepackage{amsmath,epsfig,wrapfig,graphicx,braket,bm}
\newcommand{\bfe}{\mbox{\boldmath $e$}}
\newcommand{\bfq}{\mbox{\boldmath $q$}}
\newcommand{\bfp}{\mbox{\boldmath $p$}}
\newcommand{\bfk}{\mbox{\boldmath $k$}}

\newcommand{\bfS}{\mbox{\boldmath $S$}}
\newcommand{\bfs}{\mbox{\boldmath $s$}}
\newcommand{\bfJ}{\mbox{\boldmath $J$}}
\newcommand{\bfA}{\mbox{\boldmath $A$}}
\newcommand{\bfB}{\mbox{\boldmath $B$}}

\newcommand{\calH}{{\cal H}}

\begin{document}
\title{Investigation of the Magnetic Model in Multiferroic 
NdFe$_{3}$(BO$_{3}$)$_{4}$ by Inelastic Neutron Scattering}

\author{S. Hayashida$^{1}$, M. Soda$^{1}$, S. Itoh$^{2}$, T. Yokoo$^{2}$, K. 
Ohgushi$^{3}$, D. Kawana$^{1}$, H. M. R\o nnow$^{1,4}$ and T. Masuda$^{1}$}

\affiliation{$^{1}$Neutron Science Laboratory, Institute for 
Solid State Physics, 
University of Tokyo, Tokai, Ibaraki 319-1106, Japan\\ 
$^{2}$Neutron Science 
Division, Institute of Materials Structure Science, High Energy Accelerator 
Research Organization, Tsukuba, Ibaraki 305-0801, Japan\\ 
$^{3}$Department 
of Physics, Tohoku University, Sendai, Miyagi 980-8581, Japan\\
$^{4}$Laboratory for Quantum Magnetism,
$\acute{E}$cole Polytechnique F$\acute{e}$d$\acute{e}$rale Lausanne (EPFL), 
CH-1015 Lausanne, Switzerland}

\date{\today}

\begin{abstract}
%Rare-earth ferroborates $R$Fe$_{3}$(BO$_{3}$)$_{4}$ ($R=$ rare-earth metal)
%are a series of 
%multiferroic compounds containing $R^{3+}$ and
%Fe$^{3+}$ as magnetic ions.
%The variety of the magnetic anisotropy of the $R^{3+}$ moments 
%induces diverse magnetoelectric effects 
%thorough the interaction between the Fe$^{3+}$ and $R^{3+}$ moments
%($f$-$d$ coupling).
%In these compounds the multiferroic mechanism 
%is explained by
%the spin-dependent metal-ligand hybridization model.
We performed inelastic neutron scattering measurements on single crystals of 
NdFe$_{3}$($^{11}$BO$_{3}$)$_{4}$ to explore the magnetic excitations, 
to establish the underlying Hamiltonian, and to reveal the detailed 
nature of hybridization
between the 4$f$ and 3$d$ magnetism. 
The observed spectra exhibiting a couple of 
key features, i.e., 
anti-crossing of Nd- and Fe-excitations and anisotropy gap at 
the antiferromagnetic zone center, are explained by 
the magnetic model including spin interaction in the framework of
weakly-coupled Fe$^{3+}$ chains, 
interaction between the Fe$^{3+}$ and Nd$^{3+}$ moments, and single-ion 
anisotropy derived from Nd$^{3+}$ crystal field.
%The combination of the measurements and calculations 
%reveals that the hybridization between 4$f$ and 3$d$ magnetism 
%propagates the local
%magnetic anisotropy of the Nd$^{3+}$ moment to the Fe$^{3+}$ network,
%resulting in the bulk structure of multiferroics. 
The combination of the measurements and calculations 
reveals that the hybridization between 4$f$ and 3$d$ magnetism 
propagates the local
magnetic anisotropy of the Nd$^{3+}$ moment to the Fe$^{3+}$ network,
leading to the determination of the bulk structure of both electric 
polarization
and magnetic moment in the multiferroics of the spin-dependent
metal-ligand hybridization type. 
\end{abstract}


\maketitle

\section{Introduction}
Coexistence of magnetic order and electric polarization, 
{\it multiferroicity},
has become a major topic over the past decade in condensed matter physics.
Since multiferroicity was originally discovered in perovskite 
TbMnO$_{3}$, \cite{nature426}
various multiferroic compounds have been found, including 
$R$MnO$_{3}$ ($R$ = Eu, Gd, Tb, and Dy),\cite{PRL92_RMnO3}
Ba$_{0.5}$Sr$_{1.5}$Zn$_{2}$Fe$_{12}$O$_{22}$, \cite{PRL94_BaSrZnFeO}
Ni$_{3}$V$_{2}$O$_{8}$, \cite{PRL95_Ni3V2O8}
CoCr$_{2}$O$_{4}$, \cite{PRL96_CoCr2O4} MnWO$_{4}$, \cite{PRL97_MnWO4}
CuFeO$_{2}$, \cite{PRB73_CuFeO2}
LiCu$_{2}$O$_{2}$, \cite{PRL98_LiCu2O2}
LiCuVO$_{4}$, \cite{JPSJ76_LiVCuO4} and 
Ba$_{2}$CoGe$_{2}$O$_{7}$. \cite{PRL105_Ba2CoGe2O7} 
Recent theoretical and experimental studies revealed that the 
electric polarization in these compounds is driven by 
magnetic long-range order. \cite{PRL95,PRL96,PRB73,PRB76,JPSJ76}
Since the structure of the order is determined 
by the exchange pathways and the magnetic anisotropy, 
experimental identification of the magnetic Hamiltonian 
is very important for understanding multiferroics. 

The rare-earth ferroborates $R$Fe$_{3}$(BO$_{3}$)$_{4}$ ($R$ = rare-earth 
metal) are a series of new multiferroic compounds containing $R^{3+}$ (4$f^{n}$) and 
Fe$^{3+}$ (3$d^{5}$ $S = 5/2$) as magnetic ions.
The variety of the magnetic anisotropy of the 
$R^{3+}$ moments ($R$ = Y, Pr, Nd, Sm, Gd and Tb) combined with the 
interaction between the Fe$^{3+}$ and $R^{3+}$ moments 
($f$-$d$ {\it coupling}) gives rise to diverse magnetoelectric (ME) 
effects as a function of the 
$R^{3+}$ ions.\cite{JETP81,JETP83,JETP105,JETP109,LTP36,PRB82}
In these compounds the mechanism of magnetoelectricity is explained by the 
spin-dependent metal-ligand hybridization model. \cite{PRB87,PRB89}

\begin{figure}[tbp]
\centering
\epsfig{file=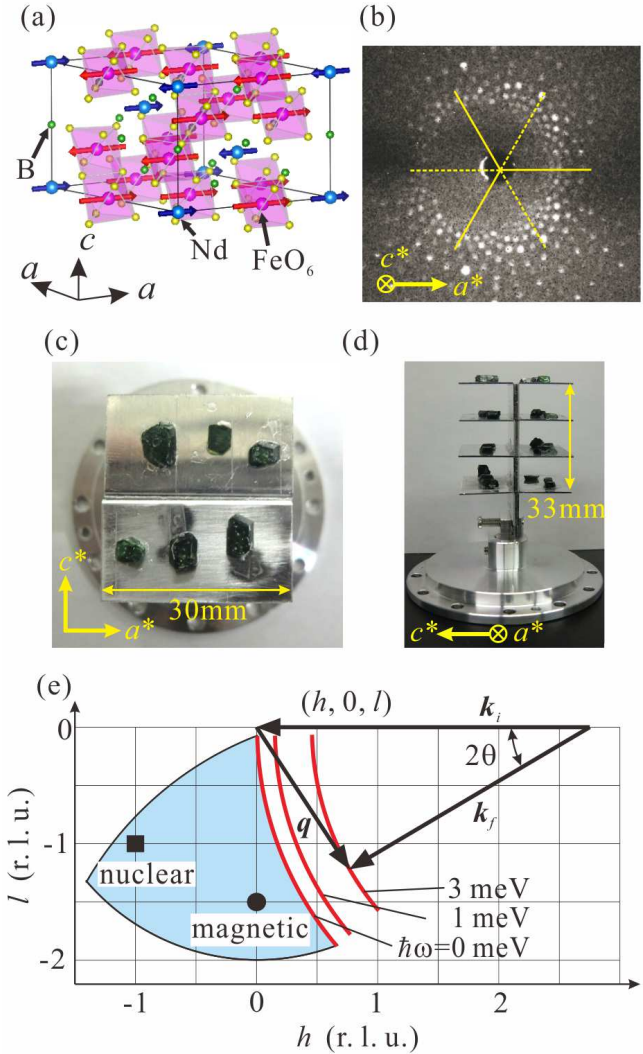,width=1.0\hsize}
\caption{(Color online) (a) The crystal structure and magnetic structure of 
NdFe$_{3}$(BO$_{3}$)$_{4}$ (hexagonal, space-group $R32$). 
(b) Laue image of a crystal with the $c^{\ast}$ - axis parallel to the 
incident X-ray beam. 
Crystals of NdFe$_{3}$($^{11}$BO$_{3}$)$_{4}$ on the alumina holder 
viewed from above (c) and from the side (d).
(e) Reciprocal space in the $a^{\ast}$ - $c^{\ast}$ plane. 
The square at $(h, k, l)=(-1, 0, -1)$ is a nuclear Bragg reflection. 
The circle at $(0, 0, -1.5)$ is a magnetic Bragg reflection. 
Red curves indicate spectra measured 
with $\bfk_{i}//a^{\ast}$ for $\hbar\omega=0,1$ and 3 meV. 
The blue shaded area indicates the range of the observed scattering plane for 
$k_{i}=2.352$ \AA $^{-1}$ and $\hbar\omega=0$ meV
when sample is rotated. }
\label{fig_1}
\end{figure}

The crystal structure has the trigonal space group $R32$, 
which belongs to the structural type of the mineral huntite 
CaMg$_{3}$(CO$_{3}$)$_{4}$. \cite{CM9} 
As shown in Fig.~\ref{fig_1}(a) 
the main feature is that distorted FeO$_{6}$ octahedra 
form spiral chains with threefold screw-axis 
symmetry along the crystallographic $c$ - axis. 
Each chain includes three Fe$^{3+}$ ions in the unit cell. 
The chains are separated by the $R^{3+}$ and B$^{3+}$ ions. 

In NdFe$_{3}$(BO$_{3}$)$_{4}$ the Nd$^{3+}$ ions (4$f^3$) carry magnetic 
moment with $J=9/2$. 
The magnetic susceptibility showed anisotropic decrease below 
29 K, and the heat capacity showed well-defined
$\lambda$ type anomaly at the same temperature, implying a phase transition
to an antiferromagnetic (AF) ordered state with N{\'e}el temperature 
of $T_N=29$ K. \cite{MMM316}
At $T \ge T_N$ the susceptibility is concave downward, 
indicating the short-range AF order because of the low dimensionality 
of the magnetic system. 
Spontaneous electric polarization simultaneously appears in the 
AF ordered phase. \cite{LTP36}
The electric polarization significantly increases 
upon applying a magnetic field 
parallel to the $a$ - axis.
The magnitude of the electric polarization reaches 
$P_{a}\sim300$ $\mu$C/m$^{2}$ at 1.3 T and 4.2 K, \cite{JETP83,JETP105}
which means that the magnetization along the $a$ - axis induces
the large electric polarization along the $a$ - axis.
A neutron diffraction study exhibited an easy-plane type 
AF order at $T \le T_N$; the Fe$^{3+}$ and 
Nd$^{3+}$ magnetic moments align ferromagnetically 
along the $a$ - axis and propagate antiferromagnetically along the $c$ - axis 
with the propagation vector $\bfk$ = (0, 0, 3/2)
in Fig.~\ref{fig_1}(a). \cite{JPCM18,PRB81}
Both of Fe$^{3+}$ and Nd$^{3+}$ moments are simultaneously ordered 
at the $T_N$, indicating non-negligible $f$-$d$ coupling. 
Further decreasing the temperature at $T \le T_{{\rm IC}}=13.5$ K, 
the commensurate (C) magnetic peak splits 
into a pair of incommensurate (IC) peaks
where the magnetic moments are in the $ab$ - plane and the AF helix propagates along 
the $c$ - axis. 
\begin{figure}[t]
\centering
\epsfig{file=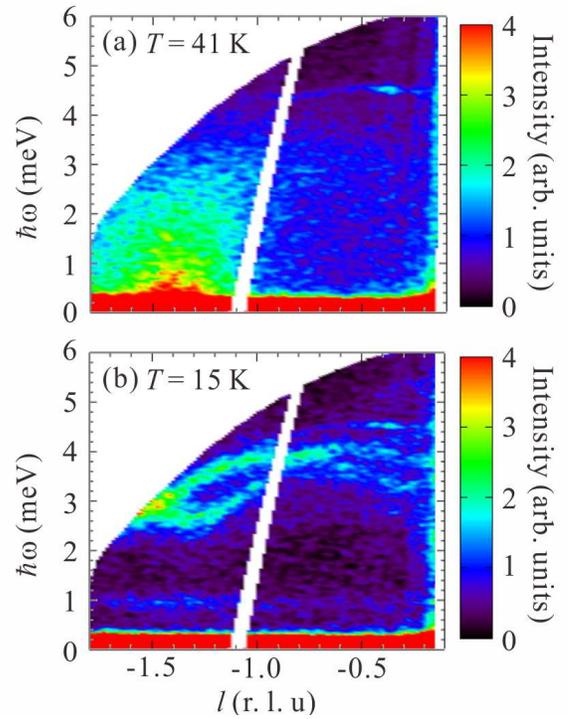,width=1.0\hsize}
\caption{(Color online) Inelastic neutron spectra projected onto $c^{\ast}$ - 
axis at (a) 41 K and (b) 15 K. The incident neutron energy was 11.46 meV.}
\label{fig_2}
\end{figure}
\begin{figure*}[t]
\centering
\epsfig{file=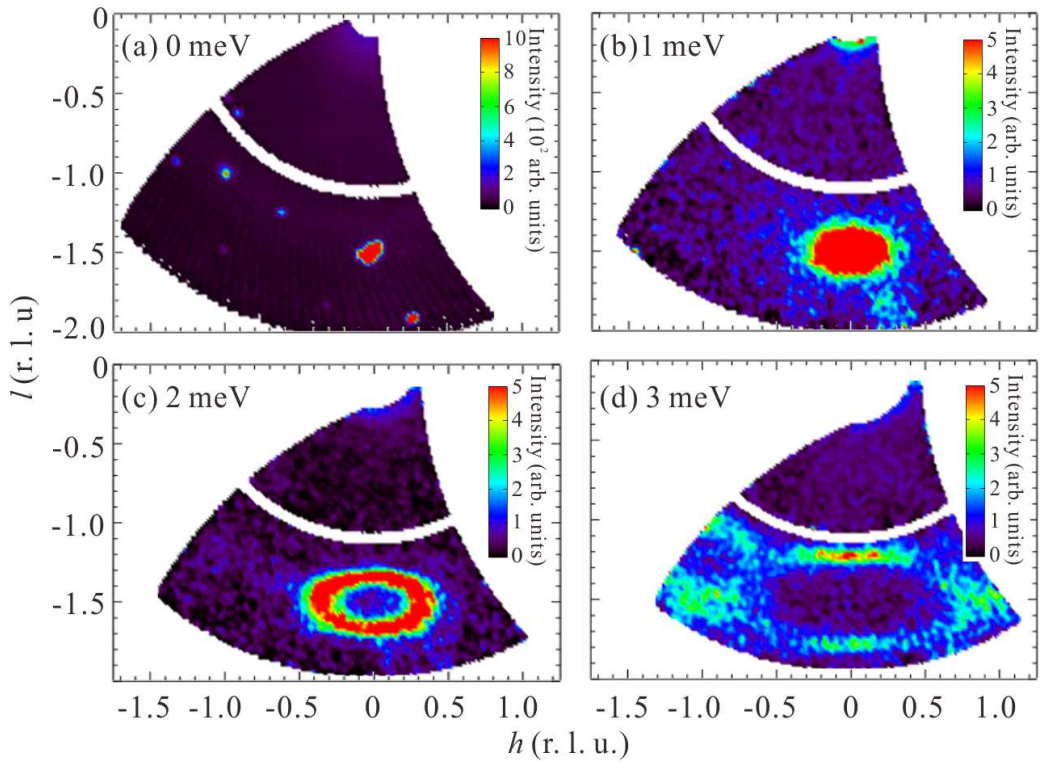,width=1.0\hsize}
\caption{(Color online) Constant energy cuts at 0 meV (a), 1 meV (b), 
2 meV (c) and 3 meV (d) in the $a^{\ast}$ - $c^{\ast}$ plane. The 
horizontal and vertical axes represent the $a^{\ast}$ and $c^{\ast}$ axes, 
respectively}
\label{fig_3}
\end{figure*}

Magnetic dynamics have been investigated by spectroscopic methods using 
electromagnetic waves.
ESR measurement detected an energy gap suggesting a uniaxial magnetic 
anisotropy in the $ab$ - plane. \cite{JETP94}
It also detected lifting of the Kramers doublet of the 
Nd$^{3+}$ ion due to the 
molecular field from the neighboring Fe$^{3+}$ ions.
Optical spectroscopy provided the energy levels 
of the crystal field of 
the Nd$^{3+}$ ion and determined the parameters of the crystal field 
Hamiltonian. \cite{PRB75}

%For the understanding of the 
%spin-driven multiferroics, it is important to 
%identify the magnetic Hamiltonian.
%Particularly 
In NdFe$_{3}$(BO$_{3}$)$_{4}$ 
exhibiting the strong $f$-$d$ coupling, 
the investigation of excitation 
spectra including Fe$^{3+}$ spin wave and Nd$^{3+}$ crystal field 
in a wide wave-vector energy space 
is crucial in order to identify the magnetic 
Hamiltonian and to unravel the detailed 
nature of hybridization between 3$d$ and 4$f$ magnetism.
Furthermore in constructing the 
Hamiltonian, careful consideration of magnetic 
anisotropy is important in the multiferroics 
of the spin-dependent 
metal-ligand hybridization mechanism type,\cite{PRB76,JPSJ76}
in which the magnetic anisotropy directly determines 
the polarization structure.

In the present paper we study inelastic neutron scattering (INS) 
spectra on NdFe$_{3}$($^{11}$BO$_{3}$)$_{4}$ 
to explore the magnetic excitations and to establish 
the underlying Hamiltonian. 
Following to the introduction 
we describe the experimental details
about the sample preparation and the setup of INS measurements
in Sec.~\ref{sec:experimental-details}.
Subsequently in Sec.~\ref{sec:experimental-results} the INS spectra 
of NdFe$_{3}$($^{11}$BO$_{3}$)$_{4}$ are demonstrated. 
We observed spin waves of the Fe$^{3+}$ moment below 6 meV
and transition between the lifted states of Kramers doublet 
of the Nd$^{3+}$ ion at 1 meV. 
A couple of characteristic features are an anti-crossing of the 
Fe- and Nd-excitations, and 
a small anisotropy gap at the antiferromagnetic zone center. 
In Sec.~\ref{sec:analysis} the magnetic model including 
an in-plane anisotropy derived from the crystal field excitation of 
the Nd$^{3+}$ moment and the non-negligible $f$-$d$ coupling is constructed. 
The observed spectra are successfully analyzed 
by the linear spin wave theory based on the model. 
The origin of the in-plane anisotropy is revealed to be 
the crystal field of the Nd$^{3+}$ ion. 
In Sec.~\ref{sec:discussion} possibility of magnetic anisotropy
of the Fe$^{3+}$ moment is discussed. 
It is turned out 
that the anisotropy is very small, 
and, instead, the Fe$^{3+}$ moment inherits an in-plane anisotropy
through hybridization with the Nd$^{3+}$ moment.
The conclusions are given in Sec.~\ref{sec:conclusion}. 
The magnetic Hamiltonian in NdFe$_{3}$($^{11}$BO$_{3}$)$_{4}$ is established 
in the present study. 
Combination of the measurement and the detailed calculation 
revealed that the hybridization between 
4$f$ and 3$d$ magnetism propagates 
the local magnetic anisotropy of the Nd$^{3+}$ ion 
to the Fe$^{3+}$ network, resulting in the bulk structure 
of multiferroics. 
%Consideration of the multiferroic mechanism reveals 
%that 
The local symmetry of the rare-earth ion is a driving force 
for the non-local multiferroicity in NdFe$_{3}$($^{11}$BO$_{3}$)$_{4}$.

\section{Experimental details}\label{sec:experimental-details}
Single crystals of NdFe$_{3}$($^{11}$BO$_{3}$)$_{4}$ were grown by 
a flux method.\cite{flux}
We first synthesized polycrystalline samples from the starting
materials, Nd$_{2}$O$_{3}$, Fe$_{2}$O$_{3}$, and $^{11}$B$_{2}$O$_{3}$. 
The stoichiometric amounts of the
starting materials with a total mass of about 16 g were mixed, ground, and put
into an alumina crucible. 
The crucible was heated at 980 $^{\circ}$C for 72 h.
The flux is Bi$_{2}$Mo$_{3}$O$_{12}$ + 3 $^{11}$B$_{2}$O$_{3}$ 
+ 3/5 Nd$_{2}$O$_{3}$; Bi$_{2}$Mo$_{3}$O$_{12}$ was synthesized by
the solid state reaction from Bi$_{2}$O$_{3}$ and MoO$_{3}$ 
inside an alumina crucible at 600 $^{\circ}$C for 24 h. 
A mixture of about 60 g 
of NdFe$_{3}$($^{11}$BO$_{3}$)$_{4}$ and the flux
with the mass ratio of 1 : 3 was put into a platinum crucible inside the
alumina crucible. The crucible was heated to 1000 $^{\circ}$C for 4 h, kept
at this temperature for 1 h, cooled to 962 $^{\circ}$C for 1 h, and slowly
cooled down to 870 $^{\circ}$C for 120 h; then the furnace was shut down to
the room temperature. The flux was removed by decanting at 900 $^{\circ}$C,
and washing the crystals with HCl solutions.

We coaligned 22 pieces of single crystals so that the crystallographic 
$a^{\ast}$ - $c^{\ast}$ plane is horizontal.
Alignment was performed by transmission Laue method using 
a high energy X-ray Laue camera.
The X-ray source was YXLON MG452 and the maximum energy of the white X-ray 
beam was 310 keV.
We recorded Laue patterns using a high-speed CCD camera, with imaging 
size 10 cm $\times$ 10 cm (1024 $\times$ 1024 pixel).
Figure~\ref{fig_1}(b) shows a Laue image of a crystal in the 
$c^{\ast}$ - plane. 
This pattern exhibits the threefold symmetry along the $c^{\ast}$ - axis.
We placed the crystals on an alumina holder as shown 
in Figs.~\ref{fig_1}(c) and \ref{fig_1}(d). 
The average mass of the crystals was 0.1 g.
The total mass of the sample was 2.1 g.

The INS experiment was performed at the High 
Resolution Chopper Spectrometer (HRC) installed in the Material and Life 
Science Experimental Facility of J-PARC. \cite{NI631,NI654,JPSJ82}
At the HRC white neutrons are monochromatized by a Fermi chopper 
synchronized with the production timing of the pulsed neutrons.
The energy transfer $\hbar\omega$ was determined from the time of flight 
(TOF) of the scattered neutrons detected at position sensitive detectors (PSDs). 
The T$_{0}$ chopper was set at 50 Hz, a collimator of 1.5$^{\circ}$ was 
installed in front of the sample, and the ``S'' Fermi chopper with 200 Hz was 
used to obtain high neutron flux.
We used a GM-type closed cycle cryostat to achieve 41 K and 15 K. 
The energy of the incident neutron beam was $E_{i}=11.46$ meV yielding an 
energy resolution of $\Delta E=0.3$ meV at the elastic position.

Figure~\ref{fig_1}(e) illustrates the $a^{\ast}$ - $c^{\ast}$ scattering plane.
Reciprocal lattice positions at $\bfq = (-1, 0, -1)$ and $(0, 0, -1.5)$ 
are the positions of nuclear 
and magnetic Bragg peaks, respectively. 
Throughout this paper $\bfq$ is expressed in reciprocal lattice unit, 
$\bfq = (h,k,l)$.  
INS spectra with $\bfk_{i}//a^{\ast}$ were measured at 41 K and 15 K.
Red curves in Fig.~\ref{fig_1}(e) indicate the measured $\bfq$-ranges for 
$\hbar\omega=0,1$ and 3 meV. 
At 15 K INS spectra that cover wide $\bfq$-range were measured 
by rotating the crystal by 70 degree in 2 degree steps.
The $\bfq$-range in the scan for $\hbar\omega=0$ meV is indicated 
by the blue shaded area in Fig.~\ref{fig_1}(e).
The range of out of plane momentum is $|q_{\perp}|<0.41$ \AA $^{-1}$.
In the following, the out of plane spectra are integrated in the central range
$|q_{\perp}|<0.13$ \AA $^{-1}$, and we show all spectra in the $(h,0,l)$ plane.
\begin{figure}[t]
\centering
\epsfig{file=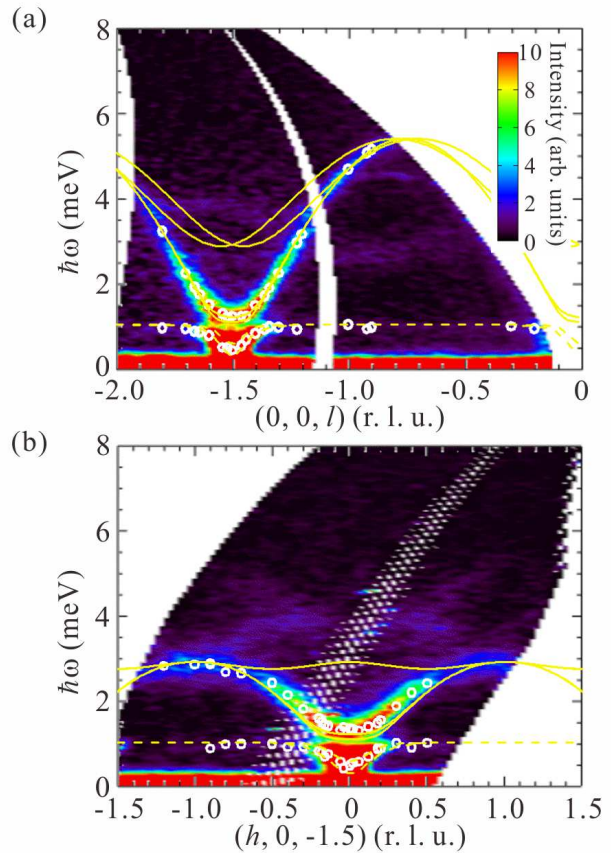,width=1.0\hsize}
\caption{(Color online) Color maps of the inelastic 
neutron scattering spectra 
obtained at HRC, along the (a) $l$ and (b) $h$ 
directions at 15 K. 
The open 
circles are the peak positions extracted Gaussian 
fits and the curves are 
the calculated spin wave dispersions. 
Fitting curves reasonably reproduce the experimental data. 
White area in the middle in the panel (a) is the gap between 
neutron detectors banks. 
}
\label{fig_4}
\end{figure}
\begin{figure}[t]
\centering
\epsfig{file=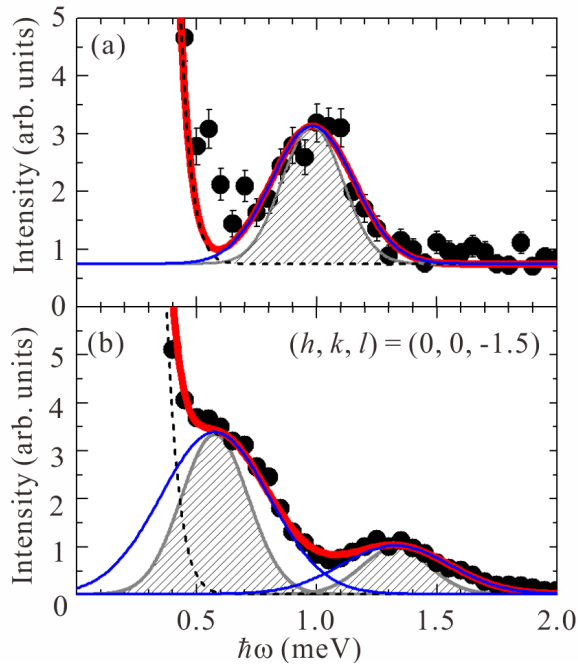,width=1.0\hsize}
\caption{(Color online) (a) $\hbar \omega$ dependence 
of the neutron intensity 
that is obtained by the integration in the $\bfq$-range of $h=[-0.5,0]$ 
and $l=[-1,-0.7]$. The solid curve is fit to the 
data by Gaussian functions. (b) 
Constant-$Q$ scan at $(0,0,-1.5)$. The spectra were integrated in the range 
of $h=[-0.1,0.1]$ and $l=[-1.6,-1.4]$. The error bars are inside the 
symbols. The solid curves are fits to the data by Gaussian functions. The gray 
shaded areas are the energy resolutions: Gaussian functions with full width at 
half maximum of 0.3 meV.}
\label{fig_5}
\end{figure}

\section{Experimental results}\label{sec:experimental-results}
The INS spectra projected onto the $c^{\ast}$ - axis 
at 41 K and 15 K are shown in Figs.~\ref{fig_2}(a) and \ref{fig_2}(b), 
respectively.
At 41 K a diffuse spectrum of paramagnetic scattering 
emerges from $l=-1.5$.
At 15 K dispersive excitations emerge in the 
energy range of 2.5 meV $<\hbar\omega<4.5$ meV, which is interpreted as 
spin waves of the Fe$^{3+}$ moments.

Figures~\ref{fig_3}(a)-\ref{fig_3}(d) display neutron spectra 
at 15 K sliced at the energies of 0, 1, 2, and 3 meV in the 
$a^{\ast}$ - $c^{\ast}$ plane. 
The white arcs are because of the absence of the neutron detectors between 
the detector banks.
In Fig.~\ref{fig_3}(a) the peak at $\bfq=(-1,0,-1)$ is 
a nuclear Bragg reflection, and the peak 
at $(0,0,-1.5)$ is a magnetic Bragg reflection.
The peaks at other $\bfq$s are not identified; they may be Bragg reflection 
from minor grains of crystals.
The rings expanding from $(0,0,-1.5)$ in Figs.~\ref{fig_3}(b)-\ref{fig_3}(d) imply that a 
dispersive excitation appears from $(0,0,-1.5)$.
The rings are flattened along the $c^{\ast}$ - axis, which means that the 
dispersion along the $c^{\ast}$ - axis is steeper than that along 
the $a^{\ast}$ - axis.
This is consistent with the naive prediction from the 
crystal structure that the 
intrachain interaction along the $c$ - axis is 
strong compared with the interchain 
interaction.

The INS spectrum at 15 K projected onto $\hbar \omega$ - $(0,0,l)$ plane 
by the integrating the neutron intensity in the ranges of 
$-0.1a^* \le  q \le 0.1 a^*$ along 
the ${\bm a^*}$ direction in the 
scattering plane and $-0.17a^* \le q \le 0.17 a^*$ 
perpendicular to the scattering plane is 
shown in Fig.~\ref{fig_4}(a). 
The spectrum projected onto $\hbar \omega$ - $(h,0,-1.5)$ plane 
by the integration in the ranges of 
$-0.17a^* \le q \le 0.17 a^*$ 
perpendicular to the scattering plane 
and $-1.6c^* \le q \le -1.4c^*$ along the ${\bm c^*}$ direction is 
shown in Fig.~\ref{fig_4}(b). 
We clearly observed the spin waves of the Fe$^{3+}$ moments around 
$\bfq=(0, 0, -1.5)$, and the flat excitation at about 
$1.0$ meV which is the 
transition between the lifted states of Kramers doublet of the Nd$^{3+}$ ion.
The spin waves of the Fe$^{3+}$ moments are more dispersive along the 
$c^{\ast}$ direction than along the $a^{\ast}$ direction, which is consistent 
with the flattened ring in Figs.~\ref{fig_3}(b)-\ref{fig_3}(d).

A series of $\hbar\omega$ dependence of the neutron 
intensity obtained by the integration in the same $\bfq$-ranges 
that were used for the display of 
Figs.~\ref{fig_4}(a) and \ref{fig_4}(b) 
were fitted by Gaussian functions to 
investigate the detailed structures of the excitations. 
The peak energies were plotted as open circles in 
Figs.~\ref{fig_4}(a) and \ref{fig_4}(b). 
These data will be used in the analysis section. 
The white circles around $\bfq = (0,0,-1.5)$ 
exhibit an anti-crossing between the spin wave of the Fe$^{3+}$ 
moments and the flat mode of the Nd$^{3+}$ moments, meaning that 
the Fe$^{3+}$ moments interact with the Nd$^{3+}$ moments.

Figure~\ref{fig_5}(a) shows the $\hbar\omega$ dependence of the neutron 
intensity obtained by the integration in the $\bfq$-range of 
$h=[-0.5,0]$ and $l=[-1,-0.7]$, where the Nd$^{3+}$-based mode 
is not affected by the Fe$^{3+}$-based spin-wave mode.
The dashed curve is incoherent elastic scattering
and the blue curve is a Gaussian fit.  
The red curve is the sum of the dashed and blue curves.
The gray shaded area is the energy resolution.
From the peak position we identify the magnitude of the energy split of 
the Kramers doublet to be 0.98 meV.
Figure~\ref{fig_5}(b) shows a constant-$\bfq$ scan at $(0,0,-1.5)$,
the AF zone center, that is 
obtained by integration in the $\bfq$-range of $h=[-0.1,0.1]$ 
and $l=[-1.6,-1.4]$.
The dashed curve is incoherent elastic scattering, 
the blue curves are Gaussian fits, 
and the red curve is the sum of the dashed and blue curves.
The gray shaded areas indicate the energy resolutions.
The magnitude of the energy gap at the AF zone center is estimated 
as $0.57$ meV.
The gap implies the existence of an anisotropy in the $ab$ - plane. 
The excitation at 1.3 meV is the transition 
between the split of Kramers 
doublet of the Nd$^{3+}$ ion.
The peak energies of the Nd$^{3+}$ ion in Fig.~\ref{fig_5}(a) 
and Fig.~\ref{fig_5}(b) are obviously different. 
The energy difference is due to the hybridization between the Fe$^{3+}$ 
and Nd$^{3+}$ modes.  
The energy widths of the excitations at $\hbar\omega=$ $0.57$ meV 
and 1.3 meV are broader than the experimental resolution. 
The energy split of $0.98$ meV and the energy gap of $0.57$ meV are 
consistent with the magnetic excitations reported from ESR measurement 
\cite{JETP94} and optical spectroscopy. \cite{PRB75}

\section{Analysis}\label{sec:analysis}
\begin{figure}[t]
\centering
\epsfig{file=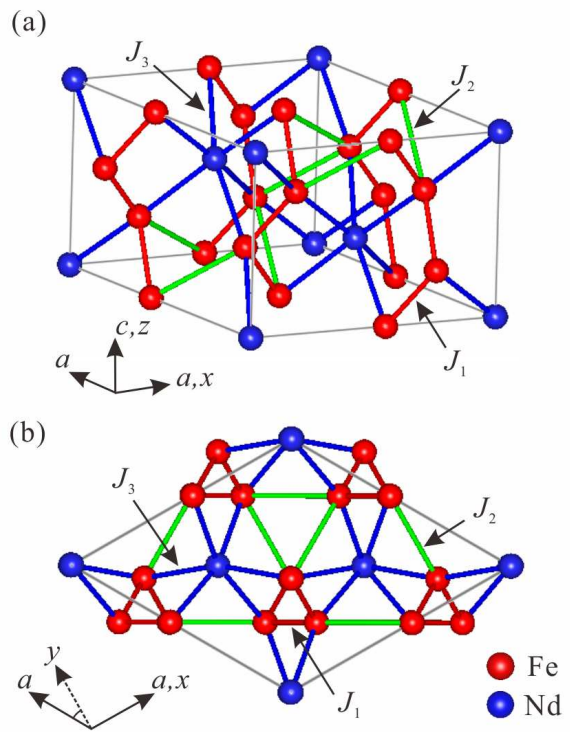,width=1.0\hsize}
\caption{(Color online) Exchange paths in NdFe$_{3}$(BO$_{3}$)$_{4}$. $J_{1}$ 
and $J_{2}$ are the nearest and 2nd nearest neighbor exchange interactions 
between the Fe$^{3+}$ moments. $J_{3}$ is the nearest neighbor exchange 
interaction between the Fe$^{3+}$ and Nd$^{3+}$ moments. The $x$ - axis is 
parallel to the $a$ - axis and the $z$ - axis is parallel to the $c$ - axis. The 
$y$ - axis is vertical to the $x$ and $z$ - axis.}
\label{fig_6}
\end{figure}
In order to identify the magnetic model realized in 
NdFe$_{3}$($^{11}$BO$_{3}$)$_{4}$, 
we consider the following Hamiltonian: 
\begin{eqnarray}
{\cal H}=-\sum_{{\rm n.n.}}J_{1} \bfS_{i}\cdot\bfS_{j}
-\sum_{{\rm n.n.n.}}J_{2} \bfS_{i}\cdot\bfS_{j} \nonumber \\
-\sum_{{\rm n.n.}}J_{3} \bfS_{i}\cdot\bfJ_{k}
+\sum_{k}{\cal H}_{{\rm CF}}(\bfJ_{k}),
\label{eq:Hamiltonian} 
\end{eqnarray}
where the $x$ - axis is parallel to the crystallographic $a$ - axis and 
the $z$ - axis is parallel to the $c$ - axis. 
$J_{1}$ and $J_{2}$ are the exchange interactions in the nearest and
2nd neighbor paths of the Fe$^{3+}$ ions as shown 
in Figs.~\ref{fig_6}(a) and \ref{fig_6}(b).
These terms mainly determine the dispersions along 
the $c^{\ast}$ - axis and 
$a^{\ast}$ - axis, respectively.
$J_{3}$ is the nearest neighbor exchange interaction between the Fe$^{3+}$ 
and Nd$^{3+}$ moments, which induces the anti-crossing between 
the Fe$^{3+}$ and Nd$^{3+}$ modes.
In Eq. (\ref{eq:Hamiltonian}), positive (negative) signs of the exchange 
parameters correspond to ferromagnetic (antiferromagnetic) exchange 
interactions.
${\cal H}_{{\rm CF}}$ is the crystal field Hamiltonian of the Nd$^{3+}$ ion.

There are five Kramers doublets in the crystal 
field of the Nd$^{3+}$ ion above 
$T_{{\rm N}}=30$ K.
The first excited energy is about 8 meV, and second one is about 17 meV 
as reported in Ref. \onlinecite{PRB75}. 
In the calculation of the low-energy excitations 
below 6 meV it is assumed that 
only the ground state in the crystal field 
of the Nd$^{3+}$ ion hybridizes with the Fe$^{3+}$ moments.
For this assumption, we introduce the ground state of the crystal field
of the Nd$^{3+}$ ion.

At the Nd$^{3+}$ site with $D_{3}$ symmetry, 
the crystal field Hamiltonian can 
be defined as follow:
\begin{eqnarray}
&&{\cal H}_{{\rm CF}}=B^{2}_{0}C^{2}_{0}+B^{4}_{0}C^{4}_{0}+
iB^{4}_{3}(C^{4}_{3}+C^{4}_{-3})\nonumber \\
&&+B^{6}_{0}C^{6}_{0}+iB^{6}_{3}(C^{6}_{3}+C^{6}_{-3})
+B^{6}_{6}(C^{6}_{6}+C^{6}_{-6}).
\end{eqnarray}
The $B^{p}_{q}$ are the crystal field parameters 
and the $C^{p}_{q}$ are the spherical tensor operators.
We used the values of the parameter $B^{p}_{q}$ reported in 
Ref. \onlinecite{PRB75}. 
Matrix elements of $J^{x}$, $J^{y}$, and $J^{z}$ in the ground state doublet
$\ket{g_{\pm}}$ are as follow:
\begin{eqnarray}
\left\{
\begin{array}{l}
\displaystyle{\bra{g_{\mp}}J^{x}\ket{g_{\pm}}= -1.67 \equiv \frac{\beta}{2},} \\
\displaystyle{\bra{g_{\mp}}J^{y}\ket{g_{\pm}}= \pm \frac{\beta}{2}i }\\
\displaystyle{\bra{g_{\pm}}J^{z}\ket{g_{\pm}}=\mp0.98 \equiv \mp\frac{\alpha}{2},} \\
\displaystyle{\bra{g_{\pm}}J^{x}\ket{g_{\pm}}=
\bra{g_{\pm}}J^{y}\ket{g_{\pm}}=\bra{g_{\mp}}J^{z}\ket{g_{\pm}}=0,}
\end{array}
\right.
\end{eqnarray}
meaning that the total angular momentum is anisotropic, favoring in-plane.
Next we inspect anisotropy within the $ab$ - plane.
The quantization axis is transformed from the $z$ - axis to the $x$ - axis
because the direction of the spin in the order state is along the $x(a)$ - axis.
Then, the matrix elements of $J^{x}$, $J^{y}$,
and $J^{z}$ in the redefined state $\ket{g_{\pm}^\prime}$ are calculated as follow:
\begin{eqnarray}
\left\{
\begin{array}{l}
\displaystyle{\bra{g_{\pm}^{\prime}}J^{x}\ket{g_{\pm}^{\prime}}
=\mp \frac{\beta}{2},} \\
\displaystyle{\bra{g_{\mp}^{\prime}}J^{y}\ket{g_{\pm}^{\prime}}
= \frac{\beta}{2},} \\
\displaystyle{\bra{g_{\mp}^{\prime}}J^{z}\ket{g_{\pm}^{\prime}}
= \pm \frac{\alpha}{2}i ,} \\
\displaystyle{\bra{g_{\mp}^{\prime}}J^{x}\ket{g_{\pm}^{\prime}}=
\bra{g_{\pm}^{\prime}}J^{y}\ket{g_{\pm}^{\prime}}
=\bra{g_{\pm}^{\prime}}J^{z}\ket{g_{\pm}^{\prime}}=0.}
\label{eq:matrix-element}
\end{array}
\right. 
\end{eqnarray}

The operator $B^{6}_{6}(C^{6}_{6}+C^{6}_{-6})$ leads
to an anisotropy within the $ab$ - plane.
In order to quantify this, we express the total angular momentum of 
the Nd$^{3+}$ ion in the ground state doublet as 
\begin{equation}
J^{x}=\frac{\beta}{2} \cos \theta, \quad
J^{y}=\frac{\beta}{2} \sin \theta,
\end{equation}
where $\theta$ is the angle between the $a$ - axis and the moment.
Thus classical energy of $B^{6}_{6}(C^{6}_{6}+C^{6}_{-6})$ becomes
\begin{equation}
B_{6}^{6}\left(C_{6}^{6}+C_{-6}^{6}\right)
=\frac{\sqrt{231}}{16}B_{6}^{6}\gamma_{J}
\left(\frac{\beta}{2}\right)^{6}\cos 6\theta ,
\label{eq:6fold}
\end{equation}
where $\gamma_{J}=-3.80\times 10^{-5}$ is a Stevens'  factor.
Eq.~(\ref{eq:6fold}) means that the classical energy of the crystal field gives
a six-fold anisotropy.
Since the sign of the $B_{6}^{6}\gamma_{J}$ is negative,
the easy-axis is the $a$ - axis that is consistent 
with the magnetic structure.\cite{PRB81}
The magnitude of the calculated anisotropy energy is $65.2$ $\mu$eV.
We effectively include the six-fold anisotropic energy as $-D \left( J^{x} \right)^{2}$
in the Hamiltonian Eq.~(\ref{eq:Hamiltonian}).
The coefficient $D$ is $23.5$ $\mu$eV, which is defined 
by the relation:
\begin{equation}
-D\left(\frac{\beta}{2}\right)^{2}
\equiv \frac{\sqrt{231}}{16}B_{6}^{6}\gamma_{J}
\left(\frac{\beta}{2}\right)^{6}\times 2=65.2 \hspace{5pt}\mu {\rm eV}.
\end{equation}

Next the operators of total angular momentum $\bfJ$ are approximated 
as the operators of pseudo-spin $s=1/2$ because 
the ground state is Kramers 
doublet and the degree of freedom is two.
The connection between operators of the total angular momentum $\bfJ$ 
and the pseudo-spin $\bfs$ is determined 
by the matrix elements Eq.~(\ref{eq:matrix-element}). 
In this approximation the operators of the total angular momentum is redefined as 
\begin{equation}
J^{x}=\beta s^{x}, \quad J^{y}=\beta s^{y}, \quad J^{z}=\alpha s^{z}.
\label{eq:angular_operator}
\end{equation}
The Hamiltonian Eq.~(\ref{eq:Hamiltonian}) is, thus, represented by
\begin{eqnarray}
{\cal H}&=&-\sum_{{\rm n.n.}}J_{1} \bfS_{i}\cdot\bfS_{j}
-\sum_{{\rm n.n.n.}}J_{2} \bfS_{i}\cdot\bfS_{j}\nonumber \\
&&-\sum_{{\rm n.n.}}J_{3} \bfS_{i}\cdot
\left(
\begin{array}{c}
\displaystyle{\beta s_{k}^{x}} \\
\displaystyle{\beta s_{k}^{y}} \\
\displaystyle{\alpha s_{k}^{z}} 
\end{array}
\right)
-\sum_{k}D \left( \beta s_{k}^{x} \right)^{2}.
\label{eq:Hamiltonian_2} 
\end{eqnarray}
We subsequently calculate the spin wave spectrum of this Hamiltonian using 
Holstein-Primakoff (HP) transformations.
The HP transformations of the spin operators of the Fe$^{3+}$ moments and 
the pseudo-spin operators of the Nd$^{3+}$ moments are written as
\begin{eqnarray}
S^{x}_{i}&=&S-a^{\dagger}_{i}a_{i},\\
S^{y}_{i}&=&\sqrt{\frac{S}{2}}\left(a_{i}^{\dagger}+a_{i}\right),\\
S^{z}_{i}&=&-\sqrt{\frac{S}{2}}i\left(a_{i}^{\dagger}-a_{i}\right),\\
s^{x}_{i}&=& s-b^{\dagger}_{i}b_{i},\\
s^{y}_{i}&=&\sqrt{\frac{s}{2}}\left(b_{i}^{\dagger}+b_{i}\right),\\
s^{z}_{i}&=&-\sqrt{\frac{s}{2}}i\left(b_{i}^{\dagger}-b_{i}\right).
\end{eqnarray}
The $a^{\dagger}_{i},a_{i},b^{\dagger}_{i}$ and $b_{i}$ are bosons operator in 
each sublattices.
The quantization axis is parallel to the $x(a)$ - axis.
We introduce spatial Fourier transformation via
\begin{equation}
c_{i}^{\dagger}(\bfq)=\frac{1}{\sqrt{N}}\sum_{\mbox{\boldmath $r$}_{i}}
c_{i}^{\dagger}e^{-i\bfq\cdot\mbox{\boldmath $r$}_{i}} \quad (i=1\sim24),
\end{equation}
where $N$ is the number of unit cells in the system, and 
$\{c_{i}\}=\{a_{i},b_{i}\}$ are the bosons operators on each sublattice.
By using this notation we obtain 
\begin{eqnarray}
{\cal H}&=&\sum_{\bfq}\sum_{ij}A_{ij}(\bfq) c_{i}^{\dagger}(\bfq)c_{j}(\bfq) \nonumber \\ 
 && + \frac{1}{2}\sum_{ij}\left[B_{ij}(\bfq) c_{i}^{\dagger}(\bfq)c_{j}^{\dagger}(\bfq) +  {\rm h.c.}\right].
\end{eqnarray}
The eigenvalues of the matrix $\left(\mbox{\boldmath $A$}+\mbox{\boldmath $B$}\right)\left(\mbox{\boldmath $A$}-\mbox{\boldmath $B$}\right)$ give the squares of the energy of the normal modes \cite{PRB56}:
\begin{equation}
\left(\bfA+\bfB\right)\left(\bfA-\bfB\right) \chi_{\tau}(\bfq)
=\left\{\hbar\omega(\bfq)\right\}^{2}\chi_{\tau}(\bfq).
\end{equation}
The dispersions obtained from this 
calculation are indicated by the yellow solid 
curves in Figs.~\ref{fig_4}(a) and \ref{fig_4}(b).
In fact we obtained 24 modes of spin waves, but due to the trigonal symmetry 
of the lattice only 8 modes have non-zero spectral weight.
The modes at the highest and second-highest energies around 
the AF zone center, ${\bfq} = (0,0,-1.5)$, 
in Fig.~\ref{fig_4}(a) are two-fold degenerated, 
and the highest energy mode in Fig.~\ref{fig_4}(b) is four-fold degenerated.
At the zone center, 
there are two modes at $\hbar \omega = 1.12$ meV and $1.23$ meV, 
and another two modes at $\hbar \omega = 0.32$ meV and $0.61$ meV.
The $0.32$ meV gap is caused by the anisotropy of the Nd$^{3+}$ moment 
in the $ab$ plane. 
It vanishes if the $D$ is set to zero.
The modes at $1.12$ meV and $1.23$  meV is the Nd$^{3+}$ 
level after hybridizing
with the dispersive Fe$^{3+}$ spin waves.
The small splittings of both modes are 
due to the easy-plane type anisotropy of the 
Nd$^{3+}$ ion, i.e., 
the effect of $\alpha \neq \beta$. 
These splittings are the origin of the observed 
broadenings of the experimental peaks 
in Fig.~\ref{fig_5}(b).
White circles in Figs.~\ref{fig_4}(a) and \ref{fig_4}(b) are 
fit by the mean energy of the split modes. 
${\chi}^2$s were calculated for the parameter set 
with the step sizes of $\delta J_1 = 0.001$ meV, 
$\delta J_2 = 0.001$ meV, and 
$\delta J_3 = 0.1$ $\mu$eV in the ranges of 
$|J_1| \le 1$ meV, $|J_2| \le 1$ meV, and 
$|J_3| \le 0.1$ meV. 
The obtained parameters set for the minimum ${\chi}^2$ is 
listed in Table~\ref{parameter}. 
The fit to the data provides excellent agreement with 
the overall spectrum. 
It should be noted that the anisotropy gap of about $0.57$ meV at the 
zone center is quantitatively reproduced by using the 
fixed parameter of $D$ = 23.5 $\mu$eV obtained from reported 
value of the parameter $B_6^6$.\cite{PRB75} 
It is revealed that the origin of the in-plane anisotropy 
is the crystal field of the Nd$^{3+}$ ion. 
\begin{table}[b]
\caption{Parameters obtained by the linear spin wave calculations}
\label{parameter}
\begin{tabular}{ccccc}
\hline 
\hline
 \quad $J_{1}$ [meV] \qquad &  $J_{2}$ [meV]  \qquad & $J_{3}$ [$\mu$eV] \qquad
  & $D$ [$\mu$eV] (fixed) \qquad  &
$\chi^{2}$ \quad  \\
\hline
\quad  $-0.482$ \qquad &  $-0.054$ \qquad & $7.9$ \qquad 
& $23.5$  \qquad & $0.993$  \quad \\
\hline \hline 
\end{tabular}\\
\end{table}\\

\section{Discussion}\label{sec:discussion}
In $R$Fe$_{3}$(BO$_{3}$)$_{4}$, the magnitude of the electric polarization
by the $R^{3+}$ and Fe$^{3+}$ ions are locally determined 
by the magnetic moments.\cite{PRB87,PRB89}
In case of NdFe$_{3}$(BO$_{3}$)$_{4}$ the existence of the in-plane anisotropy
favoring order along the $a$ - axis by the Nd$^{3+}$ and/or Fe$^{3+}$ ions
is a key to the emergence of the multiferroicity.
In the analysis section, uniaxial anisotropy of the Fe$^{3+}$ moments
is not considered for the simplicity.
In this section, we discuss possible magnetic
anisotropies in the $ab$ - plane of the Fe$^{3+}$ moments.

The conventional origin of anisotropy in Fe$^{3+}$-based 
magnets is magnetic-dipole interaction or single-ion anisotropy.
The magnetic-dipole interaction between spins $\bfS_{i}$ and $\bfS_{j}$ 
is represented by
\begin{equation}
{\cal H}_{\rm dip}=\sum_{i,j}\frac{\left(g\mu_{\rm B}\right)^{2}}{r_{ij}^{3}}
\left\{\bfS_{i}\cdot\bfS_{j}-3\left(\bfS_{i}\cdot\bfe_{ij}\right)
\left(\bfS_{j}\cdot\bfe_{ij}\right)\right\},
\end{equation}
where $r_{ij}$ and $\bfe_{ij}$ are respectively the distance and the unit vector 
along the bond between $i$ and $j$.
For a collinear in the $ab$ - plane,
the dipole-interaction energy is independent of the angle to the $a$ - axis.
This is due to the threefold screw-axis symmetry along the $c$ - axis,
which also dictates that further neighbor interactions vanish.
Thus, the magnetic-dipole interaction is not the origin of 
the anisotropy in the $ab$ - plane. 

Next, we consider the single-ion anisotropy of the Fe$^{3+}$ moments.
There is only one inequivalent site for the Fe$^{3+}$ ion and the local anisotropy 
is uniquely determined. 
Since the screw axis $3_1$ or $3_2$ is along the Fe$^{3+}$ chain, 
the FeO$_6$ octahedra are transformed by $2\pi/3$ rotation around 
the $c$ - axis one another. 
Therefore, three local coordinates, 
$\left\{X_{i},Y_{i},Z_{i}\right\}$, 
can be defined on the FeO$_6$ octahedra for the anisotropy. 
Here $\{i=1,2,3\}$ are the labels of the Fe$^{3+}$ sites. 
Since the Fe$^{3+}$ moments are collinear in the $ab$ - plane,
we discuss the anisotropy only in the $ab$ - plane.
Then, the general single-ion anisotropy to 4th order in the Fe$^{3+}$ spin-operators 
is expressed by
\begin{eqnarray}
\lefteqn{\calH_{{\rm aniso}}=\sum_{i=1,2,3}\left[a_{x^{2}}\left(S_{i}^{X_{i}}\right)^{2}
+a_{y^{2}}\left(S_{i}^{Y_{i}}\right)^{2}+a_{xy}S_{i}^{X_{i}}S_{i}^{Y_{i}}\right.} \nonumber \\
&& +a_{x^{4}}\left(S_{i}^{X_{i}}\right)^{4}
+a_{y^{4}}\left(S_{i}^{Y_{i}}\right)^{4}
+a_{x^{2}y^{2}}\left(S_{i}^{X_{i}}\right)^{2}\left(S_{i}^{Y_{i}}\right)^{2} \nonumber \\
&&\left.+a_{x^{3}y}\left(S_{i}^{X_{i}}\right)^{3}S_{i}^{Y_{i}} 
+a_{xy^{3}}S_{i}^{X_{i}}\left(S_{i}^{Y_{i}}\right)^{3}\right] .
\label{eq:single-ion}
\end{eqnarray}
The $a_{x^{2}}$, $a_{y^{2}}$, $a_{xy}$, $a_{x^{4}}$,  $a_{y^{4}}$, 
$a_{x^{2}y^{2}}$, $a_{x^{3}y}$, and $a_{xy^{3}}$ are independent coefficients.
Here we define the the local coordinate $\{X_{1},Y_{1},Z_{1}\} $
as the same as the global one $\{x,y,z\} $.
Then the relation between the spin operators defined on the 
local coordinates $\{X_{2,3},Y_{2,3},Z_{2,3}\} $
and the those defined on the global one is as follows:
\begin{eqnarray}
\left\{
\begin{array}{l}
\displaystyle{S_{2}^{X_{2}}
= -\frac{1}{2}S_{2}^{x}
+\frac{\sqrt{3}}{2}S_{2}^{y},} \\
\displaystyle{S_{2}^{Y_{2}}
= -\frac{\sqrt{3}}{2}S_{2}^{x}
-\frac{1}{2}S_{2}^{y},} \\
\displaystyle{S_{2}^{Z_{2}}= S_{2}^{z},}
\end{array}
\right. 
\label{eq:120rotation2}
\end{eqnarray}
\begin{eqnarray}
\left\{
\begin{array}{l}
\displaystyle{S_{3}^{X_{3}}
= -\frac{1}{2}S_{3}^{x}
-\frac{\sqrt{3}}{2}S_{3}^{y},} \\
\displaystyle{S_{3}^{Y_{3}}
= \frac{\sqrt{3}}{2}S_{3}^{x}
-\frac{1}{2}S_{3}^{y},} \\
\displaystyle{S_{3}^{Z_{3}}= S_{3}^{z},}
\end{array}
\right. 
\label{eq:120rotation3}
\end{eqnarray}
where the $x$ - axis is parallel to the crystallographic 
$a$ - axis and the $z$ - axis is parallel to the $c$ - axis.
These relations are substituted into Eq.~(\ref{eq:single-ion})
to express the single-ion anisotropy in global coordinates.
Hereafter we classically calculate the anisotropy energy 
of the collinear AF structure in the $ab$ - plane. 
The Fe$^{3+}$ spin operators are classically expressed by
\begin{eqnarray}
\left\{
\begin{array}{l}
\displaystyle{\left(S_{1}^{x},S_{1}^{y},S_{1}^{z}\right)=
\left( S\cos\theta, S\sin\theta,0\right),} \\
\displaystyle{\left(S_{2}^{x},S_{2}^{y},S_{2}^{z}\right)=
\left(- S\cos\theta,- S\sin\theta,0\right),} \\
\displaystyle{\left(S_{3}^{x},S_{3}^{y},S_{3}^{z}\right)=
\left( S\cos\theta, S\sin\theta,0\right),}
\end{array}
\right.
\label{eq:collinear}
\end{eqnarray}
where the Fe$^{3+}$ moments are functions of the angle $\theta$
between the $a$ - axis and the Fe$^{3+}$ moments in the $ab$ - plane. 
It is found that energy is independent on the angle $\theta$.
Consequently, the single-ion anisotropy does not give 
the anisotropy of the Fe$^{3+}$ moment in the $ab$ - plane.

In the multiferroic compound Ba$_{2}$CoGe$_{2}$O$_{7}$ with the metal-ligand 
hybridization mechanism, it was reported that an interaction between the electric 
polarization determines the magnetic 
anisotropy in the easy-plane. \cite{PRL112}
We, hence, consider the electric-polarization interaction 
in NdFe$_{3}$(BO$_{3}$)$_{4}$ using the  mechanism, 
where the local electric-polarization at the Fe$^{3+}$ site is expressed by 
$\bfp_{i}=t\sum_{l}\left(\bfe_{l}\cdot\bfS_{i}\right)^{2}\bfe_{l}$.
Here $\bfe_{l}$ is a unit vector along the bond to 
ligands (in this case oxygen), 
and $t$ is a coupling constant related with the metal-ligand hybridization and 
the spin-orbit interaction.
The polarization interaction is represented by 
$J_{{\rm p}}\sum_{ij}\bfp_{i}\cdot\bfp_{j}$.
$J_{{\rm p}}$ is ferroelectric coupling constant between the electric 
polarizations $\bfp_{i}$.
The energy of the polarization interaction 
was calculated as a function of angle 
between the $a$ - axis and the Fe$^{3+}$ moments in the $ab$ - plane.
It was, then, found that the energy is independent on the direction of 
the Fe$^{3+}$ moments in the $ab$ - plane.
Thus, the polarization interaction does not cause the uniaxial anisotropy 
in the $ab$ - plane.

Accordingly, at the Fe$^{3+}$ site the conventional sources of magnetic anisotropy 
such as the magnetic-dipole 
interaction and single-ion anisotropy, and the 
polarization interaction do not lead 
to the $a$ - axis anisotropy 
under the restriction that the 
crystal symmetry is preserved.
This means that the Fe$^{3+}$ moment does not have any uniaxial anisotropy
in the $ab$ - plane unless there is any disorder which breaks
the threefold rotation symmetry,
for instance, lattice distortion, and quantum and thermal fluctuations.\cite{JETP56,PRL73}
The direction of the Fe$^{3+}$ moment, therefore, is determined by 
the anisotropy of the Nd$^{3+}$ moment through the $f$-$d$ coupling. 
It can be said that the magnetic anisotropy of the Nd$^{3+}$ moments
by the crystal field drives the multiferroicity in NdFe$_{3}$(BO$_{3}$)$_{4}$.

The multiferroic mechanism of $R$Fe$_{3}$(BO$_{3}$)$_{4}$ is 
spin-dependent metal-ligand hybridization model\cite{PRB87,PRB89} 
where the relation between the electric polarization 
and the magnetic moment is locally determined by the symmetry of O$^{2-}$ 
ions around the magnetic ion. 
In a collinear magnetic structure 
the local magnetic anisotropy is a casting vote 
in the determination of the magnetic structure, and, consequently, 
in the determination of the electric polarization as well. 
In NdFe$_{3}$(BO$_{3}$)$_{4}$ the crystal field of the Nd$^{3+}$ ion 
is revealed to be the origin of the magnetic anisotropy, which determines the 
bulk structure of multiferroics. 
This is in contrast with the multiferroic materials of which 
the mechanism is the spin current model,\cite{PRL95,PRL96} 
where the relation is determined by the geometry of neighboring 
magnetic moments. 

\section{Conclusion}\label{sec:conclusion}
We performed INS measurements to explore the magnetic excitations, 
to establish the underlying Hamiltonian, and to 
reveal the detailed nature of hybridization between the 4$f$ and 3$d$ magnetism
in NdFe$_{3}$($^{11}$BO$_{3}$)$_{4}$. 
Overall spectra are reasonably reproduced by spin-wave calculation 
including spin interaction in the framework of weakly-couped Fe$^{3+}$ chains, 
$f$-$d$ coupling, and single-ion anisotropy derived from the Nd$^{3+}$ 
crystal field. 
Hybridization between the 4$f$ and 3$d$ 
magnetism is probed as anti-crossing of 
the Nd- and Fe-centered excitations. 
The anisotropy gap observed at the AF zone center 
is explained by the crystal field of 
the Nd$^{3+}$ ion in the quantitative level. 
Magnetic anisotropy of the Fe$^{3+}$ ion allowed in 
the present crystal structure 
is small so that it cannot be dominant. 
Combination of the measurements and calculations revealed 
that the hybridization between
4$f$ and 3$d$ magnetism propagates the local magnetic
anisotropy of the Nd$^{3+}$ ion to the Fe$^{3+}$ network, 
resulting in 
the bulk magnetic structure. 
In the multiferroics of the spin-dependent metal-ligand 
hybridization type, the local magnetic anisotropy controls 
the electric polarization, meaning that the local
symmetry of the rare-earth ion is a driving force for the
non-local multiferroicity in NdFe$_{3}$(BO$_{3}$)$_{4}$.

\section*{Acknowledgment}
We thank H. Matsuda, and K. Asoh for their contribution to the single
crystal growth.
The neutron scattering experiment was approved by the Neutron Scattering
Program Advisory Committee of IMSS, KEK (Proposal No. 2013S01 and 2014S01).
H. M. R\o nnow gratefully thank ISSP for the hospitality, 
and support from the Swiss National Science Foundation 
and its Sinergia network Mott Physics Beyond the Heisenberg Model (MPBH).
This work was supported by KAKENHI (24340077).
S. Hayashida was supported by the Japan Society for the Promotion of Science 
through the Program for Leading Graduate Schools (MERIT).

\end{document}